\documentclass[12pt]{article}

\input epsf
\begin{document}
\newcommand{\Tr}{\mbox{Tr\,}}
\newcommand{\beq}{\begin{equation}}
\newcommand{\eeq}[1]{\label{#1}\end{equation}}
\newcommand{\bea}{\begin{eqnarray}}
\newcommand{\eea}[1]{\label{#1}\end{eqnarray}}
\renewcommand{\Re}{\mbox{Re}\,}
\renewcommand{\Im}{\mbox{Im}\,}

\begin{titlepage}

\hfill hep-th/0312039

\hfill

\vspace{20pt}

\begin{center}
{\Large \textbf{BOUNDARY RIGIDITY AND HOLOGRAPHY}}
\end{center}

\vspace{6pt}

\begin{center}
\textsl{M. Porrati $^{a}$ and R. Rabadan $^{b}$}
\vspace{20pt}

\textit{$^a$ Department of Physics, New York University\\ 4 Washington Pl.,
New York NY 10012, USA}

\vspace{10pt}

\textit{$^b$  School of Natural Sciences, Institute for Advanced Studies\\
Olden Lane, Princeton NJ 08540, USA}
\end{center}
\vspace{12pt}

\begin{center}
\textbf{Abstract}
\end{center}

\vspace{4pt} {\small \noindent
We review boundary rigidity theorems assessing that, under appropriate
conditions,
Riemannian manifolds with the same spectrum of boundary geodesics are
isometric. We show how to apply these theorems to the problem of
reconstructing a $d+1$ dimensional, negative curvature space-time from
boundary data associated to two-point functions of high-dimension local
operators in a conformal field theory. We also show simple, physically
relevant examples of negative-curvature spaces that fail to satisfy in a
subtle way some of the assumptions of rigidity theorems. In those examples, we
explicitly show that the spectrum of boundary geodesics is not sufficient to
reconstruct the metric in the bulk. We also survey other reconstruction 
procedures and comment on their possible implementation in the context of the
holographic AdS/CFT duality.}
\vfill
\vskip 5.mm
 \hrule width 5.cm
\vskip 2.mm
{\small
\noindent e-mail: massimo.porrati@nyu.edu, rabadan@ias.edu}
\end{titlepage}
\tableofcontents
\section{Introduction}
The holographic principle~\cite{'t} is a potentially revolutionary new
paradigm in quantum gravity, since it gives up the idea that a
fundamental description of physics is local. In place of locality,
the principle states that the fundamental degrees of freedom that describe
quantum gravity in a region of $d+1$-dimensional space-time, called
``the bulk'' hereafter, are located
on an appropriate $d$-dimensional subspace, a ``screen'' located somewhere
in that region. A proper definition of such holographic screen can be given
also in cases where the bulk has no boundary~\cite{b}.
What is generally unknown, instead, is the
physics of the degrees of freedom that live on that screen. In the case that
the background bulk space-time is Anti de Sitter (AdS) space, much more can be
said. In that case it has been
conjectured that quantum gravity --or better string theory-- on $AdS_{d+1}$
space (times some compact manifold of dimension $9-d$) has a
dual description in terms of a $d$-dimensional
(local) conformal field theory (CFT) defined on the boundary $M_d$ of
$AdS_{d+1}$ \cite{m}.
A comprehensive review of the evidence in support of that conjecture can be
found in~\cite{agmoo}.

The relation between quantum gravity in $AdS_{d+1}$ and the CFT on $M_d$ is a
duality, because when one description is perturbative, the other is
strongly coupled. So, for instance, in the canonical case when the duality is
between the Type IIB superstring on $AdS_5\times S_5$ and N=4, $SU(N_c)$
super Yang-Mills
in four dimensions, one can trust the low-energy supergravity approximation to
the superstring in the large $N_c$ limit, and
only when the 't Hooft coupling constant of the N=4 theory,
$g^2_{YM}N$ is large.

The fact that the two dual
descriptions are never simultaneously weakly coupled
makes it difficult to establish an explicit ``dictionary'' associating states
to the
quantum gravity in AdS to states of the dual conformal field theory.
Consider in particular the case where the quantum gravity wave function
is peaked
around a given classical geometry. A natural question one can ask is how to
reconstruct this geometry from CFT boundary data only.
This question does not have as yet a complete answer, even though much progress
has been made in the last few years. For instance, proposals exist for the CFT
description of precursors~\cite{prec}, and for how to detect, through
CFT correlators, the region behind the horizon of an AdS black hole~\cite{kos}.

In this paper, we continue the program of ``holographic'' reconstruction of
space-time by looking at a special class of CFT observables, namely
the two-point correlators of  local operators with high conformal dimension.
We will investigate to
what extent they can determine the geometry of the bulk space-time.
The Green's functions we select
are particularly simple because they are directly
related to the geodesic distance of two boundary points in the (regularized)
bulk space-time.

The reconstruction of the bulk space-time from boundary data reduces,
in this approximation, to a classical problem in mathematics: the boundary
rigidity problem.
Its precise definition will be given in Section 2, here we can
formulate it as follows: {\em under what conditions are two spaces with the same
spectrum of geodesics, whose endpoints lie on the boundary, isometric?}

In Section 2, we will review the argument of ref.~\cite{br} connecting
Green's functions to geodesic distance, and we will summarize
existing theorems about boundary rigidity, paying
particular attention to the assumption necessary to prove them.
We will use some of these known results to show, for instance, 
that a small deformation of
(Euclidean) AdS space is boundary rigid in any dimension.

In Section 3, we will examine some specific examples of bulk space-times:
point particles in $AdS_3$, their equal-time sections, the BTZ black
hole~\cite{btz}, the $RP^2$ geon~\cite{geon}, and their
Euclidean continuation. We will show that some of those spaces are {\em not}
boundary rigid.
The reason for that failure will be traced back to the
violation of some of the most subtle assumptions needed in proving general
boundary rigidity theorems.
The examples of Section 3, the $AdS_3$ point particle in particular,
are in some sense the flip side of the findings in ref.~\cite{br}.

That reference used Green's functions of operators with high
conformal weight as holographic probes. Among other things, it showed
that they can detect
the formation of $AdS_3$ black holes in the collision of two point particles.
So, those simple observables are nevertheless able to detect physics
behind the black-hole horizon. In Section 3, instead, we find that there
exist situations where the bulk space-time has no horizons, yet its metric
cannot be reconstructed from the spectrum of its boundary geodesics.

In Section 4, we survey, without any pretense of completeness, other 
holographic reconstruction procedures, and we discuss which of them could be
implemented using the AdS/CFT duality, i.e. from knowledge of CFT data only. 

Section 5 contains our conclusions, together with 
a conjecture about a possible extension of boundary 
rigidity theorems, and its relation to the holographic duality.

\section{Green's Functions, Geodesics, and Boundary Rigidity Theorems}
\subsection{From Green's Functions to Geodesics}
This subsection, included here for completeness, follows closely
ref.~\cite{br}.

Near the boundary, the metric of an asymptotically Anti de Sitter space is
\beq
ds^2= {L^2\over z^2}[dz^2 + g_{\mu\nu}(z,x)dx^\mu dx^\nu], \qquad
\mu,\nu=1,.,4, \qquad g_{\mu\nu}(z,x)=g_{\mu\nu}^0(x) + O(z^2).
\eeq{m1}
All non-light-like geodesics ending on the boundary $z=0$ have infinite length,
so the space must be regularized by cutting off a small region near the
boundary, specifically, by restricting $z\geq \epsilon$.
The length $\epsilon$ has a holographic counterpart in the boundary CFT: it is
the UV cutoff one needs to regularize the theory~\cite{gkp98,w,hs}. Let us 
denote the
cutoff $d+1$ dimensional bulk with ${\cal M}_{d+1}^\epsilon$.
In ${\cal M}_{d+1}^\epsilon$, geodesics have finite length.
Moreover, in this space, the boundary-to-boundary Green's
function of a free scalar field field of mass $m$ is  well defined.
This Green's function, $G(x,y)$, with $x,y\in \partial 
{\cal M}_{d+1}^\epsilon$,
is interpreted as the (regularized) two-point function of
some scalar composite operator in the dual CFT. The conformal dimension of the
operator, $\Delta$, is (generically) the largest root of the equation
\beq
L^2m^2=\Delta(\Delta - d).
\eeq{m2}
For large mass $ mL \gg 1$,  $\Delta = mL + d/2 + O(1/mL) \approx mL$.
The Green's function of a free scalar field in AdS
can be also represented as a functional integral
\beq
G(x,y)=\int [dX(t)] \exp(-\Delta D[X]/L),
\eeq{m3}
where $D[X]$ is the length of the path $X(t)$ joining $x$ to $y$.
When $mL \gg 1$, the path integral is dominated by its saddle point, i.e.
the boundary-to-boundary geodesic joining $x$ to $y$:
\beq
G(x,y)=\mbox{const}\, \{1+ O[L/\Delta D_{min}(x,y)]\}
\exp [-\Delta D_{min}(x,y)/L].
\eeq{m4}
Notice that in Eq.~(\ref{m4}) we are ignoring inverse powers of the distance,
so, even when more than one geodesic can be drawn between the points $x,y$,
we should only take into account the contribution of the shortest one. To
include the others would be inconsistent with our approximation~\footnote{
Attempts to go beyond this limitation will be discussed in Sections 4 and 5.}.
In summary, we have found that the holographic correspondence and known
results about the semi-classical approximation to free-field Green's functions
relate, by Eq.~(\ref{m4}), a CFT quantity (the two-point function of an
operator of dimension $\Delta \gg 1$) to a geometrical quantity: the minimal
geodesic distance between the two points.

\subsection{Boundary Rigidity Theorems}
Assume that a direct problem is well behaved, i.e. that its solution
exists, is unique, stable etc. The inverse problem is
to extract some properties of the original object or system from the
solution of the direct problem. These problems in general are ill-posed (in the
sense of Hadamard): there may be no solution, or the solution may be
non-unique, or unstable (small changes in the input data may result in
large changes in the solution).
Examples of inverse problems include inverse
scattering (how to reconstruct the shape of a target, or a potential from the
scattered field at large distances), the inverse gravimetry problem, 
tomography, inverse conductivity problems, inverse seismic problems, 
many problems in inverse spectral geometry \& c.

Consider in particular 
a Riemannian manifold $({\cal M}, g)$ with a boundary. Let
$D_{min}(x,y)$ be the geodesic distance between two points at the boundary $x,
y \in \partial{\cal M}$\footnote{See reference \cite{croke02} for a survey.}.
The function $D_{min}(x,y)$ is called the hodograph (a term borrowed from
geophysics). The inverse problem is to find to what extent the Riemannian 
manifold is determined by the
lengths of the geodesics between points at the boundary. Equivalently, 
the question is: up
to what extent do the two-point functions in the conformal theory determine the
bulk metric?

Solutions to this problem come into sets, related by
diffeomorphisms that reduce to the identity at the boundary.
That, of course, changes the metric
in the interior, while keeping the same geodesic spectrum. A manifold is said
boundary rigid if there exists only one such set of solutions.

Boundary rigidity theorems analyze the uniqueness of the solution. They
give the conditions that a Riemannian manifold must satisfy to be boundary
rigid, i.e. to be completely determined by the hodograph. If we take a manifold
where there exist interior points that cannot be reached by any geodesic,
then one can always change the metric close to this point without changing the
length spectrum. So, general Riemannian manifolds are not boundary rigid. What
are the conditions that a manifold should satisfy to be boundary rigid? One of
the most natural conditions is that the manifold is simple; namely, 
its boundary is strictly convex, and every two points at the boundary are 
joined by a
unique geodesic. Such a manifold is diffeomorphic to a ball. R. Michel
conjectured in 1981 \cite{mich} that every simple manifold is rigid. Another
natural condition considered by Croke \cite{croke91} is that the manifold is
strongly geodesic minimizing. This means that every segment of a geodesic 
that lies on the
interior of the manifold is strongly minimizing, i.e. it is the unique path. 
The length spectrum determines the volume of the manifold for both 
simple and strongly geodesic minimizing manifolds.

The problem is not solved in general, but there are some partial results that
will be useful to us. Simple Riemannian manifolds with negative curvature (like
AdS) are deformation boundary rigid \cite{ps}, i.e. we cannot deform the metric
keeping the boundary distance fixed. This result was generalized \cite{s92} and
further in \cite{cds00} for compact dissipative Riemannian manifolds (convex
boundary plus a condition on the maximal geodesics) satisfying some inequality
concerning the curvature. There is a semi-global result in \cite{lsu03} when
one of the metrics is close to the Euclidean and the other satisfies a bound on
the curvature.

For general metrics, not just deformations, a theorem exists in two
dimensions~\cite{c90}: every strong geodesic minimizing manifold with
non-positive curvature is boundary rigid. This theorem has been recently
generalized to subdomains of simple manifolds in two dimensions~\cite{pu}. Any
compact sub-domain with smooth
boundary of any dimension in a constant curvature space (Euclidean space,
hyperbolic space or the open hemisphere of a round sphere) is boundary rigid
\cite{mich,gromov,bgc96}. Apart from that spaces and sub-domains of
negatively curved symmetric spaces and some products of spaces\footnote{See the
survey \cite{croke02}.} there are no other boundary rigid examples.

The Lorentzian case has not been analyzed very much. The two dimensional case
is analyzed in ref.~\cite{adh96}, which tries to extend the result of Croke
\cite{c90} to the Lorentzian case. The condition analogous to being strong
geodesically maximizing is not enough to guarantee that the manifold is
boundary rigid.
\section{Examples and ``Counterexamples''}
\subsection{Point Particle in $AdS_3$}
The metric for a point particle in $AdS_3$ is locally the same as $AdS_3$, but
with different global identifications.
It reads
\beq
ds^2= {1\over r^2+ \gamma^2}dr^2 -(r^2+\gamma^2)dt^2 + r^2 d\phi^2,
\qquad r \geq 0, \qquad 0\leq \phi < 2\pi.
\eeq{m5}
By redefining $r=\gamma \hat{r}$, $\hat{t}=\gamma t$,
$\hat{\phi}=\gamma \phi$, this metric can be recast in a standard $AdS_3$
form, but with a different periodicity for the $\hat{\phi}$ coordinate:
$0\leq \hat{\phi} < 2\pi \gamma$.  This implies of course $0\leq \gamma^2 < 1$.
Negative $\gamma^2 $ gives the non-rotating BTZ black hole metric.

To correctly analyze the geodesic between any two points at the 
boundary one has to consider the Euclidean version of the problem. 
In the Lorentzian version the problem is ill defined. The problems associated 
with Lorentzian signature (there are no geodesics between some points at the 
boundary, or an infinite number of them with the same length \& c) can already
be found in simple examples as AdS. In this section we will be mainly 
concerned with Euclidean metrics. Next we will study Euclidean $AdS_3$ 
with a point particle in three cases:  at infinite temperature, where one 
studies its constant time section (which is the same as in the Lorentzian 
problem), at zero temperature, and finally at finite, nonzero temperature.
We will find that, in some cases, non-rigidity appears.

\subsubsection{Constant Time Section}
Consider now the $t=0$ section of this metric. Its geodesics can be easily
found, e.g. using the Hamilton-Jacobi method. A standard calculation gives
the angular distance between the boundary
endpoints of the geodesic, $\Delta \phi$,
as a function of its minimum distance from the center, $\bar{r}$:
\beq
\Delta \phi= {1\over \gamma} \theta,
\qquad \cot \theta= {\bar{r}^2 -\gamma^2
\over 2\gamma \bar{r}}.
\eeq{m6}
(Here we chose $\gamma >0$).
By definition, $0\leq \theta < \pi$. In this range, Eq.~(\ref{m6}) is
one-to-one. This does not mean that there is only one geodesic joining any two
boundary points! Indeed,
when the angular distance $\Delta\phi$ is in the range $\pi < \Delta\phi
< \pi/\gamma$, we have a second geodesic joining the same two
boundary points, with $\Delta \phi'=2\pi-\Delta\phi < \Delta\phi$.
Since Eq.~(\ref{m6}) is one-to-one, this means that the minimum radii of the
two geodesics are different, hence the geodesics are distinct.

So, even if our space is ``almost'' $AdS_3$, and its sectional curvature is
negative, this space may not be boundary rigid, since it fails to satisfy the
simplicity condition. Moreover, it is singular at $r=0$; removing the point
$r=0$ makes the space non-simply connected, so, again, non-simple.

If we were given the lengths of {\em all} geodesics between boundary points,
it would be still far from obvious that the point-particle space could
be deformed without changing {\em some} geodesic lengths.
In our case, though, more than one geodesics can be drawn
between the same two points, so we have to be careful about the
identification of physically meaningful holographic data.

As we mentioned in Subsection 2.1, the physical quantities one is given in the
boundary theory are the two-point function of composite operators. Geodesic are
used to obtain a saddle point approximation of these functions. Since
the saddle point approximation neglects inverse powers of the geodesic
distance [see Eq.~(\ref{m4})], one should also neglect contributions from
sub-dominant saddle points.
So, the physical data are the lengths of {\em minimal} geodesics in between
boundary points. Generically speaking, the
minimal geodesic spectrum is  not enough to reconstruct the bulk metric from
boundary data.
In our case, one can be more specific,
and prove that there exist deformations of
the metric that do not change the spectrum of minimal length geodesics.
So, not only the conditions for boundary rigidity are not met in our simple
example, but we can explicitly show that the bulk metric can be changed without
affecting boundary data.

To see this, notice that the shortest geodesic is that for which
$\Delta\phi<\pi$.
This means that no minimal-length geodesic can come closer to the center than
\beq
r_{min}=\min_{0\leq \theta\leq \gamma\pi}\bar{r}= \gamma \sqrt{{
1-\cos (\gamma\pi/2) \over 1+ \cos (\gamma\pi/2) }}.
\eeq{m7}
So, any change of the metric confined to the region $r<r_{min}$ is
undetectable, within our approximation.

Now, let us ask whether it is possible to smooth out the singularity at $r=0$
without changing the spectrum of minimum-length geodesics. This would mean
that holography could be blind to qualitative features of the bulk space
geometry, such as the very existence of singularities.
It is convenient to change coordinates in Eq.~(\ref{m5}) by setting
$r=\gamma\sinh \rho$, and write the metric at $t=0$ as
\beq
ds^2=d\rho^2 + \gamma^2\sinh^2 \rho d\phi^2, \qquad \rho>0.
\eeq{m8}
Eq.~(\ref{m7}) implies that the minimum distance $\rho_{min}$ probed by
minimal-length geodesics obeys $\gamma\sinh \rho_{min} <1$.
Now the question is,
can we smooth out the metric by changing only the region $\rho<\rho_{min}$,
while preserving some basic characteristics of the metric, for instance,
that the curvature is negative?
The answer is {\em no}. To see this, consider the change
$\gamma\sinh \rho \rightarrow F(\rho)\geq 0$. The new range of the
coordinate $\rho$ is from $\rho_0$, the point where $F$ vanishes, to $+\infty$.
Smoothness at $\rho_0$ requires $(dF/d\rho)|_{\rho_0}=1$.
To leave the metric outside $\rho_{min}$ unchanged, we must also require
$ (dF/d\rho)|_{\rho_{min}}=\gamma \cosh \rho_{min} $

To keep the curvature negative, we must have $d^2 F/d\rho^2 > 0$, whence the
inequality
\beq
 (dF/d\rho)|_{\rho_{min}} - (dF/d\rho)|_{\rho_0} = \gamma \cosh \rho_{min}-1=
\int_{\rho_0}^{\rho_{min}} {d^2 F \over d\rho^2} d\rho > 0 .
\eeq{m9}
By using the value of $r_{min}=\gamma\sinh\rho_{min}$ given in Eq.~(\ref{m7}),
we finally find that, in order to smooth out the singularity without changing
the geodesic spectrum, we must have
\beq
\gamma \cosh \rho_{min}= \gamma \sqrt{{
2 \over 1+ \cos (\gamma\pi/2) }} > 1.
\eeq{m10}
This equation is never satisfied in the range $0 < \gamma <1$.

\vspace{0.6cm}
So we have seen that there is no metric preserving rotational invariance that
coincide with the point particle metric in the region accessible by geodesics
and that has negative curvature. That means that all the metrics with this
hodograph have positive curvature in some region so the theorems about
dispersive manifolds with negative curvature (ref. \cite{ps} e.g.)
do not apply. We can extend this proof for general
deformations of the metric by considering the integral
\beq
k(\Sigma) = \frac{1}{4 \pi} \int_{\Sigma} {\cal R},
\eeq{m11}
where $\Sigma$ is a region in the interior of the Euclidean section of the
space. On a compact manifold without boundary, $k$ is the Euler number.
In two dimensions the scalar-curvature density is a total derivative.

In our case, the curvature has two contributions: one from the point particle
(a delta function at its position) and one from the AdS space itself. The first
contribution to the number $k$ can be expressed in terms of the deficit angle
$\delta = 2 \pi (1 - \gamma)$
\beq
k_{part}(\Sigma) = \frac{1}{2 \pi} \delta.
\eeq{m12}
The AdS space has constant negative curvature ${\cal R} = -2/R^2$.
So, the total
contribution in a region that contains the point particle is:
\beq
k(\Sigma) = \frac{1}{2 \pi} \delta - \frac{1}{2\pi R^2} Vol_{\Sigma}.
\eeq{m13}
The metric Eq.~(\ref{m8}) has $R=1$, so the critical volume when $k=0$ is:

\beq
Vol_c = {R^2} \delta = 2\pi (1-\gamma).
\eeq{m14}

By Eq.~(\ref{m10}),
the volume of the ball $B$ of radius $r_{min}$ --i.e. the region not probed by
minimal-length geodesics-- is
\beq
V_B= 2\pi \gamma \left( \sqrt{{
2 \over 1+ \cos (\gamma\pi/2) }} -1\right).
\eeq{m15}
Can we  change the metric within a region $\Sigma_0 \subset B$ to a {\em
smooth}, negative-curvature one, without touching the outside metric? Again,
the answer is no, since if this were possible, then, for that metric, 
$k(\Sigma_0)<0$. On the other hand, for the AdS point-particle metric:
\beq
k(\Sigma_0)= (1-\gamma) - \frac{1}{2\pi R^2} Vol_{\Sigma}>
(1-\gamma) - \gamma \left( \sqrt{{
2 \over 1+ \cos (\gamma\pi/2) }} -1\right)>0.
\eeq{m16}
So we need $k(\Sigma_0)$ to be positive for a metric, and negative for 
another. This is impossible, because for any open region $\Sigma$,
$k(\Sigma)$ is invariant under any change of
the metric inside $\Sigma$, that reduces to the identity on its boundary,
since the scalar curvature is a total derivative.

\subsubsection{Finite and Zero Temperature $AdS_3$ with a Point Particle}
Now let us consider the whole Euclidean $AdS_3$ with a point particle in it. 
It is easy to see that the shortest geodesic joining points separated by a 
very long Euclidean time can get arbitrarily close to the origin, 
as the time interval gets larger. Let us consider geodesics with only a time 
separation ($\Delta \phi =0$).
The trajectory satisfy the equation:

\beq
\frac{dr}{dt} = \frac{(\gamma^2+r^2)}{E} \sqrt{m^2 (\gamma^2+r^2) - E^2}, 
\eeq{rr1}
that can be integrated to:
\beq
t = \frac{\gamma}{2}  \log{\frac{\gamma \sqrt{m^2 (\gamma^2+r^2) - E^2} + E r}
{\gamma \sqrt{m^2 (\gamma^2+r^2) - E^2} - Er}},
\eeq{rr2}

where $E$ and $m$ are the energy and the mass of the particle.
 
\begin{figure}
\centering \epsfxsize=2in \hspace*{0in}\vspace*{.2in}
\epsffile{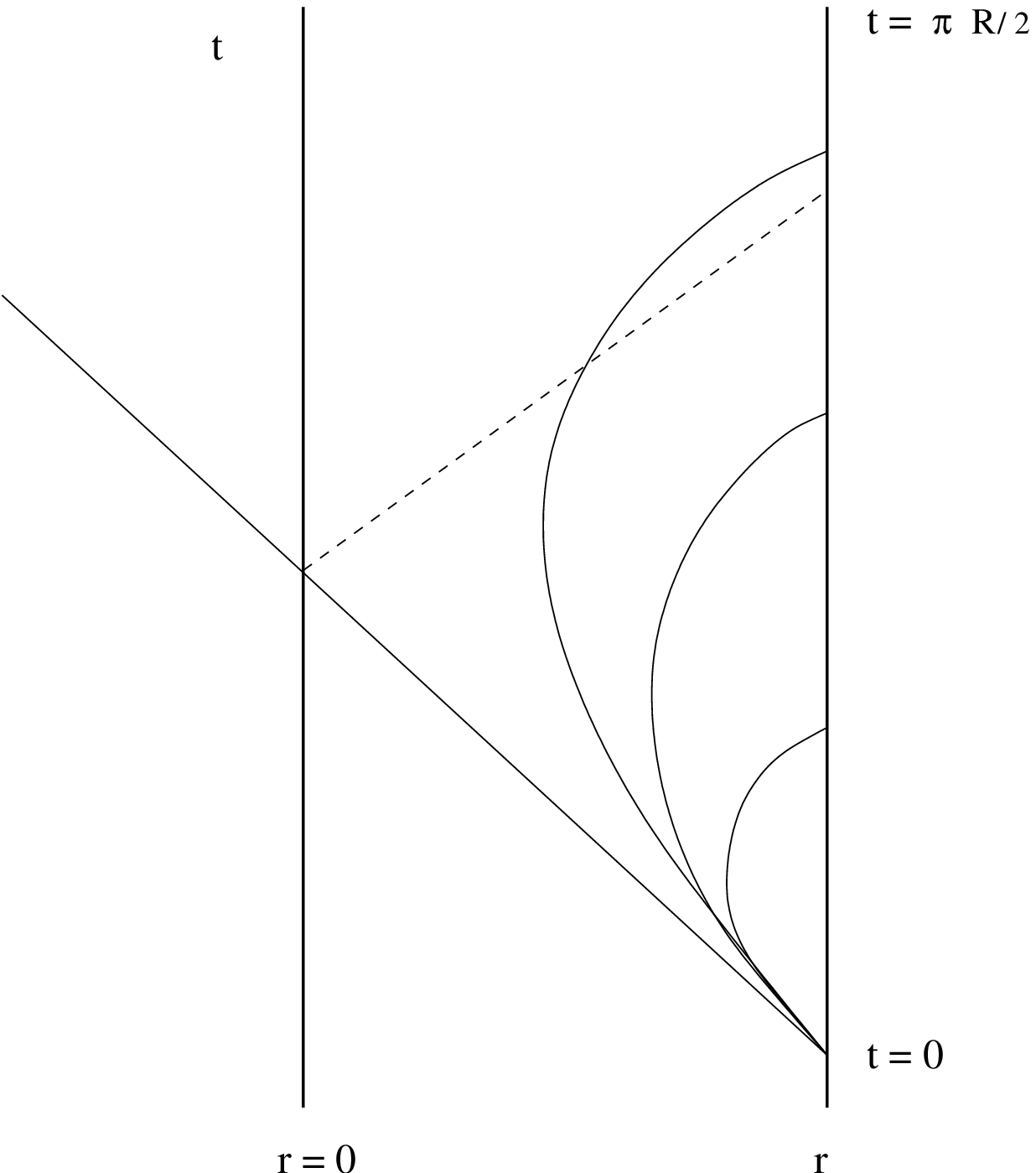}
\epsfxsize=3in
\epsffile{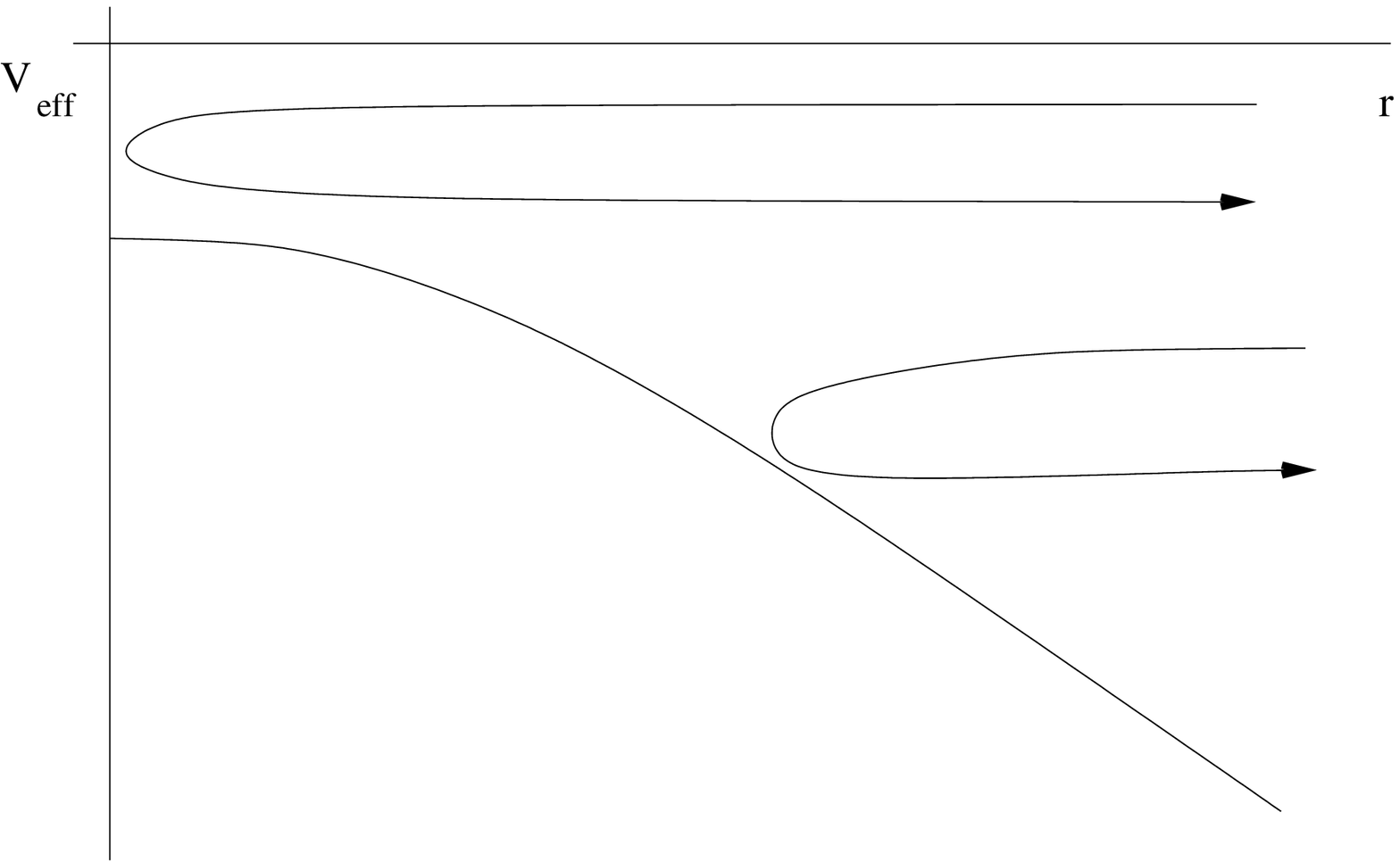}   
\caption{\small  Left: geodesics without angular momentum in Euclidean 
AdS$_3$. Right: effective potential for particles in Euclidean AdS$_3$  
without angular momentum.} 
\label{euclideanads2}
\end{figure}


Starting at the boundary there is a family of solutions with 
$|E^2| > |m^2|\gamma^2$ that do not touch the origin (see 
figure~\ref{euclideanads2}). The geodesics start from the boundary and go 
back at $\Delta \phi = 0$. The time to come back is:

\beq
\Delta T = \gamma \log{ \frac{E + m \gamma}{E - m \gamma} }.
\eeq{rr3}

Notice that for $|E^2| \rightarrow |m^2|\gamma^2$ the time interval diverges. 
That means that they can be arbitrarily long, and joining any two points on
the boundary. The turning point is at $r_c^2 = E^2/m^2 - \gamma^2$. 
For $|E^2| \leq |m^2|\gamma^2$ the geodesics pass through the origin and 
reach the antipodal point $\Delta \phi = \pi$. But, as we have seen from 
the constant time sections, these are not shortest geodesics.

So in the whole $AdS_3$ with a point particle the shortest geodesics cover 
the whole space (except the point where the point particle is located) due 
to long time geodesics.

\vspace{0.5cm}
The finite temperature case can be obtained by imposing the 
periodicity conditions $t \rightarrow t + \beta$. 
In this case the shortest geodesics cannot cover the whole space, 
as there is a maximum to the time difference  $\Delta T = \beta /2$. So, 
there is a region close to the point particle that cannot be reached by the 
shortest geodesics: the higher the temperature the larger the region. 
Notice that one needs both a point particle in the AdS space and 
finite temperature to have manifest non-rigidity.

\subsection{Lorentzian BTZ Black Hole}
\subsubsection{Description of the Space}
One can represent the non-rotating BTZ black hole as an orbifold of $AdS_3$
by a boost~\footnote{In this subsection we will follow
references~\cite{btz,kos,hvk}.}.
To define the action of the boost, let us define $AdS_3$ as a hyperboloid
in a flat space of signature $(+, +. -, -)$:
$x_0^2 + x_1^2 - x_2^2 - x_3^2 = 1$.
The boost action is:
\beq
x_1 \pm x_2 \rightarrow e^{\pm 2 \pi r_+} (x_1 \pm x_2).
\eeq{r1}
There is a line of singularities at the fixed points of the boost
$x_1 = x_2 = 0$.
Near the singularity, the metric reduces to that of a Milne universe
times a line.

Let us decompose $AdS_3$ into three regions, each of which is further
subdivided into four others, classified by a pair of signs $\eta_{1,2} = \pm$:
\begin{itemize}
\item Region 1: $x_1^2 - x^2_2 \geq 0$, $x_0^2 - x^2_3 \leq 0$,
\bea
x_1 \pm x_2 & =  &\eta_1 \frac{r}{r_+} e^{\pm r_+ \phi} \nonumber \\
x_3 \pm x_0 & =  &\eta_2 \frac{\sqrt{r^2 - r_+^2}}{r_+} e^{\pm r_+ t}.
\eea{r2}
\item Region 2: $x_1^2 - x^2_2 \geq 0$, $x_0^2 - x^2_3 \geq 0$,
\bea
x_1 \pm x_2 & = &\eta_1 \frac{r}{r_+} e^{\pm r_+ \phi} \nonumber \\
x_3 \pm x_0 & = &\eta_2 \frac{\sqrt{r_+^2 - r^2}}{r_+} e^{\pm r_+ t}.
\eea{r3}
\item Region 3: $x_1^2 - x^2_2 \leq 0$, $x_0^2 - x^2_3 \geq 0$,
\bea
x_1 \pm x_2 & = &\eta_1 \frac{\sqrt{r^2 - r_+^2}}{r_+} e^{\pm r_+ t} \nonumber
\\
x_3 \pm x_0 & = &\eta_2 \frac{r}{r_+} e^{\pm r_+ \phi}.
\eea{r4}
\end{itemize}

Notice that regions 1 and 3 reach the boundary at
$r \rightarrow \infty$, while in region 2 the radial variable ranges
from 0 to $r_+$. The singularity, located at the fixed points of the orbifold
action, is at $r = 0$, i.e. in region 2.

The boost identifies $\phi$ with $\phi + 2\pi$ in regions 1 and 2, and
$t \rightarrow t + 2\pi$ in region 3.

The metric in these new coordinates is:
\beq
ds^2 = - (r^2 - r_+^2) dt^2 + \frac{dr^2}{(r^2 - r_+^2)} + r^2 d\phi^2,
\eeq{r5}
then, the coordinate $t$ is timelike in regions 1 and 3, 
and spacelike in region 2. That gives closed timelike curves in region 3. 
To go from region 2 to region
3 we have to pass very close to the singularity. We will try to find results
that are independent of the eventual resolution of the singularity in the 
final, complete theory of quantum gravity,
so, we will not study geodesics that come close to it, and 
consider only regions 1 and 2.

The boundary is made of a set of disconnected patches labeled by the region
where they belong: $1_{(\eta_1,\eta_2)}$ and $3_{(\eta_1,\eta_2)}$.

\subsubsection{Boundary Geodesics}

Let us take a spacelike geodesic starting at the boundary of region $1_{++}$.
To begin with, we do not 
want to consider geodesics that get infinitely close to the
singularity. Then, we have to restrict ourselves
to geodesics ending at the boundary of
$1_{++}$ (that do not cross the horizon) or at the boundary of $1_{+-}$
(that cross the horizon).

We use the Hamilton-Jacobi method to obtain the time and angular difference
between the two points at the boundary in function of the energy and
angular momentum:
\beq
\Delta t = 2 \int_{r_c}^{\infty} \frac{dr}{N(r)^2}
\frac{E}{\sqrt{E^2 - V(r)^2}} ,
\eeq{r6}
and
\beq
\Delta \phi = 2 \int_{r_c}^{\infty} dr \frac{l}{r^2}
\frac{1}{\sqrt{E^2 - V(r)^2}} ,
\eeq{r7}

where $N(r)^2 = r^2 - r_+^2$ and $V(r)^2 = N(r)^2  (l^2/r^2 -1)$
is the effective potential.

The above integrals can be explicitly solved by
\beq
\Delta \phi = \frac{1}{r_+} \log\left( \frac{(E^2 - (l+r_+)^2)r_c^2}{r_c^2
(E^2 - l^2 - r_+^2) + 2 r_+^2 l^2}\right),
\eeq{r8}
and
\beq
\Delta T = \frac{1}{r_+}
\log\left(
\frac{4 r_+^2 l^2 - (E^2 -l^2 -r_+^2)^2 +(E^2 -l^2 +r_+^2)
\sqrt{\Delta}}{2 (r_c^2 - r_+^2)(2 E r_+ + E^2 -l^2 + r_+^2)}
\right),
\eeq{r9}
where $r^2_c = \frac{1}{2} (l^2 + r_+^2 -E^2 + \sqrt{\Delta})$ is the positive
root of $E = V(r)$ and $\Delta = (l^2 + r_+^2 -E^2)^2 - 4 r_+^2 l^2$.

\begin{figure}
\centering \epsfxsize=4in \hspace*{0in}\vspace*{.2in}
\epsffile{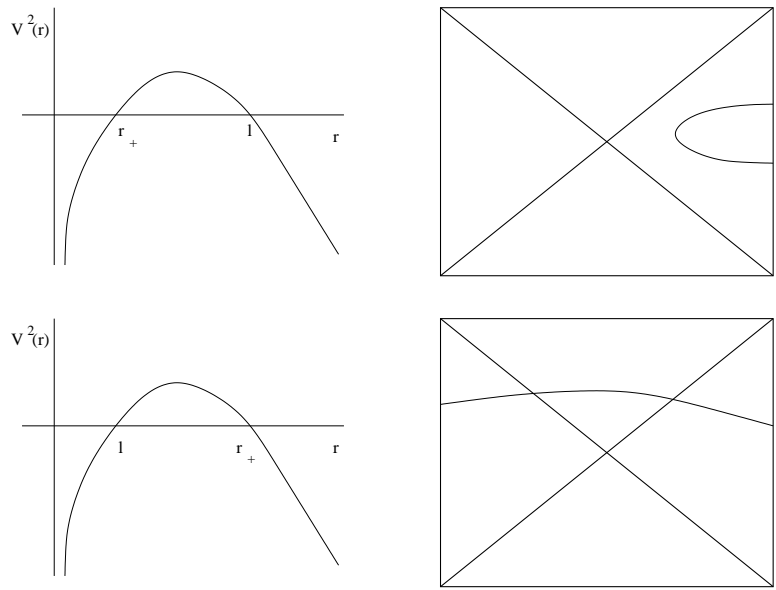}
\caption{\small Potential and corresponding trajectories on the Poincar\'e
diagram for trajectories that do not approach the singularity
[$0 < E^2 < (l - r_+)^2$]. For $r_+ < l$, the geodesic does not reach the
horizon and goes back to the boundary $1_{++}$ (upper left picture).
For $r_+ > l$, the geodesic reaches the horizon and arrives at the boundary
$1_{+-}$ (lower left picture).}
\label{BTZ}
\end{figure}

For energies $0 < E^2 < (l - r_+)^2$ the particle does not reach
the singularity and is reflected to the boundary. For larger energies,
$E^2 > (l - r_+)^2$, the particle crosses the horizon and reaches the
singularity. To stay away from the singularity, we will restrict to energies
$0 < E^2 < (l - r_+)^2$. In this range of energies,
we have two possible behaviors: some geodesics are reflected back to the
boundary of region $1_{++}$ (if $r_+ < l$), while other geodesics cross the
horizon and reach the boundary of region $1_{+-}$ (if $l< r_+$)
(see figure~\ref{BTZ}).

At $E=0$, the particle arrives at the point $r = r_+$ (if $r_+ > l$) or
$r = l$ (if $r_+ < l$). It never crosses the horizon.
In the case $r_+ > l$ it reaches the other boundary, and it is reflected
back when $r_+ < l$. When we increase the energy, while keeping $l$ fixed,
both $\Delta T$ and $\Delta \phi$ increase and become
infinity when the energy reaches its maximum.

\subsubsection{Geodesics Outside the Horizon}

Let us consider first the case $r_+ < l$. One can get arbitrarily
close to the horizon with geodesics both of whose ends belong to the boundary
by taking $l \rightarrow r_+$ and $E \rightarrow 0$. In this limit
$\Delta \phi$ diverges as we are getting close to the horizon. That means
that the geodesics wind many times close to the horizon.

That can be easily seen for Euclidean sections, i.e $E=0$,
and for geodesics within the region $1_{++}$. The angular variable can be
expressed in terms of the angular momentum
\beq
e^{r_+ \Delta \phi} = \left( \frac{l + r_+}{l - r_+} \right)^2.
\eeq{r10}
The shortest geodesics can probe the space up to a minimum
radius $r_c = r_+/\tanh(\pi r_+/4)$. Now, if we increase the energy,
the geodesics with fixed angular difference are farther away from the horizon.
This means that $r_c$ is a boundary beyond which no shortest geodesic (with
both ends at the same boundary) can penetrate.

\vspace{0.5cm}
Let us take the endpoints at $y_i = (t_i, r= 1/\epsilon, \phi_i)$ for very
large $r$ and let us assume that $(\Delta \phi + 2 \pi n)^2 > (\Delta t)^2$
for all integers. Then, there exist infinitely many geodesics connecting
these two points, labeled by an integer number $n$.
The proper length of the geodesics is:
\beq
\sinh^2(L/2) = \frac{1}{\epsilon^2 r_+^2} \sinh^2[r_+ (\Delta\phi)/2] -
\left( \frac{1}{\epsilon^2 r_+^2} - 1 \right) \sinh^2[r_+ (\Delta t)/2].
\eeq{r11}

When the regulator $\epsilon$ is taken to zero we get:
\beq
\exp(L_n) = \frac{2}{\epsilon^2 r_+^2} \left[ \cosh^2(r_+ (\Delta\phi +
2 \pi n)) - \sinh^2(r_+ (\Delta t)) \right].
\eeq{r12}

The behavior of these geodesics is easy to understand: they can wind several
times close to the horizon. The closer they get to the horizon,
the higher is their winding number $n$.

\subsubsection{Geodesics Crossing the Horizon}

Consider now the case when the particle crosses the horizon, i.e.
$l < r_+$. Notice that by taking $T \rightarrow T + i \pi/r_+$ we pass
from region $1_{++}$ to region $1_{+-}$.

To illustrate how the geodesics behave,
let us take $l = 0$.
In both cases $\Delta \phi = 0$. By naively continuing the integrals we find:
\beq
r _+ \Delta T =
\log\left(
\frac{|E^2 -r_+^2|}{(E + r_+)^2}
\right) + \pi i.
\eeq{r13}
This is interpreted as a geodesic going from region $1_{++}$ to region
$1_{+-}$.

Now, we can ask ourselves if for these geodesics there exists a region that
cannot be explored by shortest geodesics (the first question here is what we
mean by ``shortest''). As the geodesics probe inside the horizon,
that region has to lie in the interior of the horizon.
Similarly to the case when geodesics end on the same
boundary, we start with $E = 0$ ($\Delta T = 0$). The geodesic touches
the horizon and reaches the other boundary. When we increase the energy
for fixed angle, the angular momentum decreases (i.e. the orbit get closer to
the center). To go very close to the singularity we have to take $l=0$,
that is the case just explained above.

\subsection{Euclidean BTZ Black Hole}

The metric of the Euclidean BTZ black hole is defined starting
from the Lorentzian one by continuing to imaginary time:
\beq
ds^2 = (r^2 - r_+^2) d\tau^2 + \frac{dr^2}{(r^2 - r_+^2)} + r^2 d\phi^2,
\eeq{r14}
where the radial variable goes from $r = r_+$ to infinity and the
Euclidean time has period $\pi/r_+$ to avoid conical singularities at $r=r_+$.
The boundary is a two dimensional torus parametrized by the Euclidean time
and the angular variable. The interior is a solid 3-d torus, where the
points $r= r_+$ form a circle at its center.
The manifold is non simple, as there is more than one geodesic between any
two points at the boundary (geodesics can wind). That is easy to see by
considering the uncompactified version, obtained by taking the angular
variable $\phi$ non periodic. This space is again the Euclidean AdS (that is
best seen by defining a new variable $x^2 = r^2 -r_+^2$), and, as we know,
this space is boundary rigid.  Notice that this example can also be
interpreted as AdS at finite temperature, where the same kind of reasoning
applies. The temperature is identified with $T=r_+/\pi$.

The shortest geodesics reach every point at the interior. This is easily
seen because the sections of constant $\phi$ are disks, were the shortest
geodesics can connect every two points.
This is a necessary but not sufficient condition for boundary rigidity.
Whether the space is actually boundary rigid is a nontrivial question, which
is answered in the affirmative if the conjecture in Section 5 is true.

\subsection{The $RP^2$ Geon}

The $RP^2$ geon \cite{geon} can be obtained from $AdS_3$ by quotienting by
the action of a discrete group generated by:

\beq
x_1 \pm x_2 \rightarrow e^{\pm \pi r_+} (x_1 \pm x_2).
\eeq{r15}

and $x_3 \rightarrow -x_3$. So, this space is the quotient of the BTZ black
hole by a $Z_2$ symmetry.

In region $1$ of the BTZ black hole, the geon corresponds to identifying
under the transformation $\phi \rightarrow \phi + \pi$,
$\eta_2 \rightarrow - \eta_2$ and $t \rightarrow -t$. That is,
region $1_{++}$ is mapped into region $1_{+-}$, i.e. the two exterior regions
of the BTZ black hole are interchanged. The geon is thus a black hole with a
single exterior region, isometric to the region $1_{++}$ of the BTZ black hole.
Spacelike hypersurfaces are quotients of a cylinder (parametrized by
$x \sim r-r_+$ and $\phi$) by a freely acting
$Z_2$: $\phi \rightarrow \phi + \pi$ and $x \rightarrow -x$.
Topologically, this is $RP_2$ minus the point at $r = \infty$.

In region 2, the action of the $Z_2$ group is the same as in region 1.
In particular, it interchanges region $2_{++}$ with region $2_{+-}$. The
Penrose diagram is half of the Penrose diagram of the BTZ black hole, with
the upper left part reflected into the lower right part, and the lower left
part into the upper right one.

In region $1$ of the BTZ black hole, the geon corresponds to identifying
$t$ with $t + \pi$, $\eta_2$ with  $ - \eta_2$, and $\phi$ with $ -\phi$.
As in the BTZ black hole, there exist closed timelike curves.

\vspace{0.5cm}
The geodesics ending at the boundary are as in the BTZ black hole,
taking into account that the  boundary of $1_{++}$ is equivalent to the
boundary of region $1_{+-}$. So, for every two points at the boundary,
there are geodesics that cross the horizon.

\subsection{Euclidean $RP^2$ Geon}

To construct the Euclidean $RP^2$ geon one takes the Euclidean BTZ, and
quotients it by a $Z_2$ symmetry: $\phi \rightarrow \phi + \pi$ and
$t \rightarrow -t$. The sections at fixed radius are Klein bottles
(the two dimensional torus with a $Z_2$ identifications).
At the horizon, the Klein bottle degenerates to a circle.

The geodesics connecting points at the boundary can be easily computed by
considering the Euclidean BTZ geodesics, and identifying the points at the
boundary as above. The geodesics of the BTZ black hole can be obtained from
the Euclidean $AdS_3$ ones by the identification $\phi \rightarrow \phi +
2 \pi$. Said in another way, the geodesics of the geon are the same as in
$AdS_3$ when taking into into account the identification between
boundary points: $\phi \rightarrow \phi + \pi$ and $t \rightarrow -t$.

\begin{figure}
\centering \epsfxsize=3in \hspace*{0in}\vspace*{.2in}
\epsffile{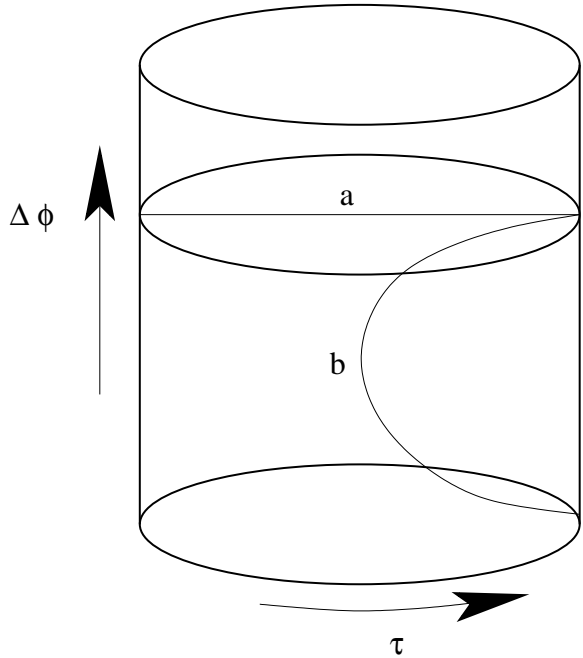}
\caption{\small Two candidates for the shortest geodesic in the geon.
Geodesics of type $a$ cover the whole space, while geodesics of type $b$ do
not reach a portion of the space close to the horizon. }
\label{geon}
\end{figure}

Let us write the BTZ black hole metric in the form:
\beq
ds^2 = r^2 d\tau^2 + \frac{dr^2}{1+r^2} + (1+r^2) d\phi^2,
\eeq{r16}
with periods $\tau \rightarrow \tau + 2 \pi$ and $\phi \rightarrow \phi +
2 \pi r_+$. To see if the shortest geodesics cover the whole space we have to
compare the distance between two geodesics (see figure \ref{geon}): the first
one, that is also present in the BTZ, has $\Delta T = \pi$ and $\Delta \phi =
0$ [let us call it geodesic $a$, joining point $( \tau = \pi/2, \phi)$ to
$(\tau = 3 \pi/2, \phi)$]. The other one, that joins points identified by the
$Z_2$ symmetry, has $\Delta T = 0$ and $\Delta \phi = \pi r_+$
[geodesic $b$, joining point $(\tau = \pi/2, \phi)$ to  $(\tau = \pi/2, \phi +
\pi r_+)$, that is the $Z_2$ image of $(\tau = 3 \pi/2, \phi)$].

The difference in length of the two geodesics is:
\beq
l_a - l_b = \log\left[ \frac{4 e^{- \pi r_+}}{(1 - e^{- \pi r_+})^2} \right].
\eeq{r17}

For small $r_+$ ($r_+ \ll 1$), the geodesic $b$ is the shortest one. It
never reaches the center. It is easy to see that in this case there is a
region inside the geon that cannot be reached by any shortest geodesics.

For large $r_+$ ($r_+ \gg 1$), the geodesic $a$ is the shortest.
It lies on a constant $\phi$ section of the torus, and, as in the BTZ black
hole, covers the whole space.

In the $RP^2$ case, one can construct a linear combination of two-point
correlation functions that does not receive contributions from the shortest
geodesic~\cite{geon}. So, both geodesics $a$ and $b$ can be unambiguously
determined by boundary data. Equivalently, in this case, the boundary data are
the lengths of all geodesics that lift to minimal-length ones in the $Z_2$
cover of the $RP^2$ geon, that is the BTZ black hole. They do probe
the entire space.
So, while one cannot reconstruct the geon metric from the spectrum of
shortest geodesics only, more refined boundary data may allow to establish a
rigidity theorem.

\subsection{Higher-Dimensional Finite-Temperature AdS Black Holes}

Now, let us consider the higher-dimensional AdS Black Hole. In this case, 
there are two geometries contributing the boundary $S^1 \times S^{d-1}$: 
the AdS Schwarzschild Black Hole ($X_2 = R^2 \times S^{d-1}$) and AdS at 
finite temperature ($X_1 = S^1 \times R^{d}$). Unlike in dimension three,
the Euclidean version of these spaces have different topology. In \cite{hp,w},
it has been shown that the dominant geometry at low temperatures is AdS at 
finite temperature, while at high temperature the dominant contribution is the
 black hole. As we have seen from the previous examples, constant-time 
geodesics in finite-temperature AdS cover the whole space, so we expect this 
space to be boundary rigid. At higher temperature, the dominant contribution 
comes from the black hole that, unlike in three dimension, 
is not boundary rigid. 
That can be seen using constant-angle geodesics (then the problem is reduced 
to a disk parametrized by time and the radial coordinate).

This is a very interesting case, since the same boundary admits two different 
geometric theories, one that is boundary rigid and other that is not. 
The different geometries have been identified in \cite{w} with two different 
phases of the same boundary CFT.

\begin{table}[htb] \footnotesize
\begin{center}
\begin{tabular}{|c|c||c|c|c|}
\hline 
Space & Description & Boundary Rigid  & Non  & Boundary Rigid  \\
      &             &                 & Boundary Rigid & Cover \\
\hline\hline 
Point Particle in AdS$_3$ & Constant Time Section &    & X  & N/A \\
\hline
               & Zero Temperature      & X   &    & X \\
\hline
               & Finite Temperature      &    &  X  & X \\
\hline
\hline
BTZ Black Hole &  $AdS_3/Z$   &  X  &    &  X \\
\hline
\hline
$RP^2$ Geon $r_+\ll 1$   & $AdS_3/Z_2 \otimes_S Z$   &    &  X  & X \\
\hline
$RP^2$ Geon $r_+\gg 1$   & $AdS_3/Z_2 \otimes_S Z$  & X  &     & X \\
\hline
\hline
Finite Temperature $AdS_d$  &  $AdS_d/Z$ & X   &    & X\\
\hline 
$AdS_d$ Black Hole       & & & X & N/A \\
\hline 
\end{tabular}
\end{center} 
\caption{ Summary table of the different examples analysed in this 
section. 
\label{table} }
\end{table}

\section{Other Bulk Reconstruction Procedures}
We have seen that the leading contribution of massive particle propagators
reproduces a unique metric when the manifold is boundary rigid. We have seen
several familiar examples of three dimensional manifolds that are boundary
rigid and other manifolds that are not.

We may ask about other structures that can be obtained from the field theory
that will tell us other information about the Riemannian manifold. Here we
review other procedures that will allow us to partially reconstruct the
interior Riemannian manifold.

\subsection{Dirichlet-to-Neumann Map}

Let $({\cal M},g)$ be a Riemannian manifold with boundary $\partial{\cal M}$,
and consider the following problem: find the field $\phi$
such that $\Delta_g \phi = 0$ on ${\cal M}$ with a given boundary value
$\phi|_{\partial{\cal M}} = J$, where $J$ is a source at the boundary.
The Dirichlet-to-Neumann map is the value of the normal derivative of the
solution to the above problem at the boundary:
$\partial_n \phi|_{\partial{\cal M}} = n^i \partial_i
\phi|_{\partial{\cal M}}$. In this way we can define a unique function
depending on the sources $\partial_n \phi|_{\partial{\cal M}}(J)$.

The map is directly related to field theory observables. In the AdS/CFT 
correspondence, the metric $g$ has a double pole at the boundary 
$\partial {\cal M}$ [see Eq.~(\ref{m1})].
This requires that we regularize the manifold by cutting it off at finite
proper distance from the boundary, $z=\epsilon$, as we did in Subsection 2.1. 
In the dual interpretation in
terms of CFT data, $1/\epsilon$ is a UV cutoff of the field theory, needed 
to properly define composite operators. The field $\phi$ is dual to a CFT 
operator, $O$. When the linearized equation of motion of $\phi$ is 
$\Delta_g \phi = 0$, then the operator $O$ has conformal dimension 
$\Delta=d=\mbox{dim}\,{\partial{\cal M}} $. 
The field $\phi(z,x)$ can be expanded as~\cite{w,hs,hs2}
\beq
\phi(\epsilon,x)=\phi_0(x) + \epsilon^2\phi_2(x) +...+ \epsilon^d\log 
\epsilon^2 \phi_d(x) + \epsilon^d \psi_d(\epsilon,x). 
\eeq{m18}
The coefficients $\phi_2,..\phi_d$ are {\em known, local} functions of 
$\phi_0(x)$ and $\psi_d(\epsilon,x)=\psi_d(0,x) + O(\epsilon^2)$. So, in
Eq.~(\ref{m18}) there are two unknown functions: $\phi_0(x)$ and 
$\psi_d(\epsilon,x)$. The Dirichlet-to-Neumann data allow to fix them both.
In the limit $\epsilon\rightarrow 0$, $\phi_0(0)$ is identified with 
the source $I$ of the the operator $O$, while $\psi_d(0,x)$ becomes 
proportional to the VEV of the operator $O$. More precisely~\cite{w,hs,hs2},
\beq
\psi_d(0,x)=4\langle 0 | O(x) \exp\left( -\int_{\partial{\cal M}}IO\right) 
|0\rangle .
\eeq{m19}

The inverse problem\footnote{This inverse problem has appeared in several
fields. It was proposed by Calderon \cite{c90} in 1980, motivated by
geophysical prospection. It also appears in Electrical Impedance Tomography
(EIT) in trying to obtain the conductivity of a medium by making voltage and
currents measurements on the boundary.} consists in extracting information
about the Riemannian manifold from the Dirichlet-to-Neumann map $\partial_n
\phi|_{\partial{\cal M}}(J)$.

It is conjectured that the for manifolds of dimension  $dim({\cal M})>2$ the
Dirichlet-to-Neumann map determines the Riemannian manifold uniquely (for
dimension two it determines uniquely the conformal class of the metric). It has
been proved by Uhlmann and collaborators \cite{leu,lu,ltu,u} that the
Dirichlet-to-Neumann map determines the Riemannian metric for real-analytic
manifolds of dimension $dim({\cal M})>2$ and the conformal structure for
$C^{\infty}$ manifolds and $dim({\cal M}) = 2$.

\subsection{Scattering Relation}

Imagine a geodesic that starts and ends at the boundary of a Riemannian
manifold. The scattering relation is a function that, for a starting point and
initial velocity of a geodesic at the boundary, gives the final point and
the final velocity at the boundary: $\alpha(x_i,v_i) = (x_f,v_f)$. For that
map to be well
defined, we demand that the Riemannian manifold is non-trapping, i.e. that
each maximal geodesic is finite. The scattering relation is an involution
($\alpha^2$ is the identity).

In the dual field theory, one may think of obtaining this relation from a
two-point correlator of a dimension-$\Delta\gg 1$ operator as follows.

The correlator of two (bare) operators in the regularized theory with cutoff 
$\epsilon$ is, thanks to Eq.~(\ref{m4}),
\beq
\langle O(x) O(y)\rangle_\epsilon\propto \exp [-\Delta D_{min}(x,y)/L].
\eeq{m20}
We can convolute $\langle O(x) O(y)\rangle_\epsilon$ with a function $f(x)$,
localized around $x_i$ within an uncertainty $\delta$:
\beq
\int_{\partial {\cal M}^\epsilon} f(x)\langle O(x) O(y)\rangle_\epsilon.
\eeq{m21}
This function
also localizes the momentum components along the boundary, 
$p=-i\partial/\partial x$, within an uncertainty
$1/\delta$ around a central value $p_i$. 
In the geodesic approximation, the mass-shell condition 
$L^2\sum_{m=1}^{d+1}p_mp^m=\Delta(\Delta-d)=m^2L^2$ holds. 
So, we also know the 
normal component of the momentum, within an uncertainty $1/\delta$.
Whenever $p_i\gg 1/\delta$, and the boundary is sufficiently smooth, we may
reasonably approximate the result of the convolution by assigning an
initial position $x_i$ and initial velocity $v_i=p_i/m$ to the
geodesic  
\beq
\int_{\partial {\cal M}^\epsilon} f(x)\langle O(x) O(y)\rangle_\epsilon
\approx \exp [-\Delta D_{v_i}(x_i,y)/L].
\eeq{m22}
Clearly, in this approximation, the two-point function is nonzero only for a
specific value of $y$, to wit: the final point $x_f$. 
Eq.~(\ref{m22}) also uniquely defines the final velocity $v_f$ (by 
convoluting it with an approximate eigenstate of the final momentum) 
hence the scattering relation.

The problem with this procedure is that it does not give an exact dispersion
relation, but only an approximate one. This is because Eq.~(\ref{m22}) is 
exact only in the classical limit. Even within the semiclassical 
approximation, Eq.~(\ref{m22}) is contaminated by extremal trajectories 
beginning near $(x_i,v_i)$. To be concrete, imagine the case that two
trajectories join the points $x_i,x_f$; one with initial velocity $v_i$, the 
other with initial velocity $w_i$. Both trajectories contribute to 
Eq.~(\ref{m22}). To estimate the contribution of the second, denote by
$\tilde{f}$ the Fourier transform of $f$.
For $f$ Gaussian of width 
$\delta$ we have, approximately, 
$\tilde{f}(mv)\approx \exp[-\delta \Delta^2(v-v_i)^2/2L^2]$; so,
Eq.~(\ref{m22}) becomes, with obvious notations
\beq
\int_{\partial {\cal M}^\epsilon} f(x)\langle O(x) O(y)\rangle_\epsilon
\approx \exp [-\Delta D_{v_i}(x_i,y)/L] +  
\exp [-\delta \Delta^2 (w_i-v_i)^2/2L^2 - \Delta D_{w_i}(x_i,y)/L].
\eeq{m22a}
The second contribution can be neglected only if 
\beq
\exp[-\delta \Delta^2 (w_i-v_i)^2/2L^2-\Delta D_{w_i}(x_i,y)/L+
\Delta D_{v_i}(x_i,y)/L]\ll 1. 
\eeq{m22b}
This restricts the validity of Eq.~(\ref{m22}) to
manfolds which, even though non-simple, do not have geodesics with lenght 
too close to the minimizing one. To make precise statements on 
non-minimizing geodesics, we need additional information on the CFT, as 
explained later in Subsection 4.4 and in the Conclusions.

The inverse problem is whether the scattering relation determines the metric.
In the case that the manifold is simple the scattering relation is equivalent
to the boundary distance function for the two points at the boundary
\cite{mich}. It has been shown in \cite{pu} that in two dimensional simple
manifolds the Dirichlet-to-Neumann map is determined by the scattering
relation. So in this case the scattering relation, the hodograph and the
Dirichlet-to-Neumann map are related.

\subsection{Bulk to Boundary Functions}

A complete information about the metric on the manifold is given by the bulk to
boundary Green function for very massive fields. Again, in the limit of very
high mass this function is very well approximated by the distance $r_x(y)$
between a point at the interior of the manifold $x \in {\cal M}$ and a point at
the boundary $y \in \partial{\cal M}$. Now let us consider the function ${\cal
R}$ that assigns to every point $x \in {\cal M}$ its boundary distance function
${\cal R}: x \in {\cal M} \rightarrow r_r \in L^{\infty}(\partial{\cal M})$,
where $L^{\infty}(\partial{\cal M})$ is the space with the norm:
\beq
|| r || = sup_{z \in (\partial{\cal M})} |r(z)|.
\eeq{r51}

Let us call ${\cal R}({\cal M})$ the set of all boundary distance functions. In
\cite{kkl} it is shown that one can construct a differential structure and a
metric on the set of the boundary distance functions such that it becomes
isometric to the original Riemannian manifold.

To see how it works, let us consider first the case of geodesically regular
manifolds (there is a unique geodesic between any two points in the bulk, and
the geodesic goes to the boundary). Then we will consider the general case.
Take two points $x$ and $x'$ in the interior of ${\cal M}$ and compute the
function:
\bea
f :  \partial{\cal M} & \rightarrow & R^+, \nonumber \\
        y    & \mapsto & |r_x(y) - r_{x'}(y)|.
\eea{r52}

Using the triangular inequality one can easily see that $d(x,x')
\geq f(y)$ for all points $y$. If in addition the manifold is regular,
there is a unique geodesic that joins the points $x$ and $x'$ and
that goes till the boundary (let us call this point at the boundary $y_c$). At
this point the inequality is saturated: $d(x,x') \geq f(y)$. As there is only
one geodesic (regular manifold) that means that the
distance between the two points is just the  maximum of $f(y)$.

That is: we can read what is the distance between any two points inside from
the bulk to boundary function.

A direct extension of this reasoning will be to consider the case when
there are several geodesics between the points  $x$ and $x'$ (i.e one is the
shortest and the other wind around the manifold).  Then there are several local
maxima. The lower of these maxima is the distance (measured by the shortest
geodesic). The only requirement is that the geodesics arrive to the boundary.
The proof of how the metric is reconstructed from the bulk to boundary
functions for general manifolds can be found in \cite{kkl}.

So, if we know the bulk-to-boundary distance, we can easily reconstruct the 
bulk metric. Unfortunately, the holographic interpretation of 
this quantity is rather mysterious. In specific theories, as 
$SU(N)$, ${\rm N}=4$ super Yang-Mills, one may be able to extract it from 
expectation values of, say, $\Tr F_{\mu\nu}F^{\mu\nu}$, computed on a 
one-instanton background~\cite{bk1,bgkr}. 
The actual implementation of this program on a generic manifold is still 
unclear to us.
\subsection{Spectral Boundary Data}

Now, let us consider a different type of data. They are obtained from a
differential operator (it must be elliptic, so we must work in Euclidean space)
of the form:
\beq
D = - \frac{1}{ \sqrt{g}}\partial_i( \sqrt{g} g^{ij} \partial_j) + V
\eeq{r53}
where $V$ is an arbitrary functions on ${\cal M}$. This operator can be
obtained from an action:
\beq
S = \int f \sqrt{g} \left[(\partial \phi)^2 + V \phi^2\right]
\eeq{r54}
that can be interpreted as the action of 
a massive particle $\phi$ with a position-dependent
``mass'' $m^2 = V$. Notice that if we have a dimensional reduction of the 
form ${\cal M} \times {\cal M'}$ to ${\cal M}$ with a warped metric, 
the warp factor can always be interpreted as a modification of the potential.

Now, let us consider the Dirichlet problem on ${\cal M}$, i.e.
$\phi|_{\partial{\cal M}} = 0$. The boundary spectral data is the collection of
all the eigenvalues $\lambda_k$ and the normal derivatives of the
eigenfunctions at the boundary $\partial_n \phi_k|_{\partial{\cal M}} = n^i
\partial_i \phi_k|_{\partial{\cal M}}$.

In \cite{bk,kkl} it is show how the spectral data determines uniquely the
manifold ${\cal M}$, the metric $g$ and the variable mass $V$. It is shown that
the boundary spectral data determines the set of boundary distance functions.
As we have shown in the previous paragraph this also defines the metric.

To obtain these data from a boundary CFT, we need additional assumptions,
either on the bulk manifold or on the analytic structure of the CFT.
For instance, if the space-time manifold has a time-like global Killing 
vector, then we can reinterpret ${\cal M}$ as its constant-time section. Then
the eigenvalues $\lambda_k$ are determined by the conformal weights of the
CFT~\cite{w}. 

For a generic bulk manifold, this interpretation is not possible. 
Nevertheless, spectral boundary data can be obtained if the two-point
correlator of CFT operators of arbitrary dimension $\Delta$, 
$F(\Delta,x,y)\equiv\langle O^\Delta (x) O^\Delta (y)\rangle$, is a known
{\em analytic} function of $\Delta$. In this case, the poles of this function
determine the $\lambda_k$. Of course, analyticity in $\Delta$ is a rather tall
order on a generic CFT!

\section{Summary, Conclusions, Speculations}
In Section 3 we found that the Euclidean geon is in some cases non-rigid, 
yet its metric can be determined if we know all its boundary geodesics, not
only the minimizing (shortest) ones. Indeed, in all examples we gave, 
manifest non-rigidity was associated to the
existence of a region unreachable by {\em shortest} geodesics. That
limitation was crucial. Longer geodesics, with nonzero winding number, can
reach all points inside all spaces studied in Section 3. So, neither
the $AdS_3$ point particle at finite temperature, nor its $t=0$ section, nor
the small ($r_+\ll 1$) geon possess regions that cannot be reached by some
geodesic. 
If one can find an unambiguous way to determine the length of
non-minimal geodesics from boundary data, then these spaces may be boundary
rigid after all. 
They all share one common property: they are quotients by discrete
isometries of a boundary rigid space: $AdS_3$. This leads us to the following
conjecture:
\begin{quotation}
\noindent
{\em Quotients of boundary rigid manifolds by discrete isometries are also
boundary rigid if they have the same scattering relation.}
\end{quotation}

In stating the conjecture, we used the fact that the natural way to obtain 
the spectrum of {\em all} boundary geodesics is through the scattering 
relation. 

If true, this conjecture would give a concrete, computationally effective way
to reconstruct a bulk metric from simple holographic data.

So, it is important to see if the scattering relation can be determined by the
CFT correlators. As we saw in Subsection 4.2, the ``physical'' way of 
obtaining it is only approximate. To do better, we must assume some additional 
analyticity property in $\Delta$ for the two-point correlators of the CFT. 
Specifically, if $F(\Delta,x,y)$, defined in
the previous subsection, is analytic for $\Delta\gg 1$, then one can us the
geodesic approximation to arrive at
\beq
F(\Delta,x,y)=\sum_i \mbox{const}_i\, \{1+ O[L/\Delta D_{i}(x,y)]\}
\exp [-\Delta D_{i}(x,y)/L].
\eeq{m23}
Here the sum extends to {\em all} geodesics between the boundary points $x$ 
and $y$. Since $F(\Delta,x,y)$ is analytic in
$\Delta$, its inverse Laplace transform, $\tilde{F}(t,x,y)$ 
contains delta functions located precisely at $t=\Delta_i(x,y)$
\beq
\tilde{F}(t,x,y)= \sum_i \mbox{const}_i\,\delta[t-\Delta_i(x,y)] + ....
\eeq{m24}
The ellipsis denote less singular terms.

Finally, we must remember that in the case when the CFT is a gauge theory, 
there are additional non-local 
observables with a simple geometric interpretation
in the holographic dual. One such observable is the Wilson loop. 
In particular, the correlator of two Wilson loops is $\propto\exp(-S_m)$, where
$S_m$ is the minimal surface between the two loops~\cite{loops}. This 
leads
to another inverse problem; namely: when is it possible 
to reconstruct the metric of a manifold with a known spectrum of minimal 
surfaces in between boundary loops?

\subsection*{Acknowledgments}
We would like to thank D. Berenstein, M. Kleban, J. Maldacena, and especially
G. Uhlmann for helpful discussions. 
The work of M.P. is supported in part by NSF through 
grants PHY-0070787 and PHY-0245068. R.R. is supported by DOE under 
grant DE-FG02-90ER40542.

\end{document}